\documentclass[12pt,preprint]{aastex}



\slugcomment{\today}
\shorttitle{Spitzer Observation of Lyman Break Galaxies}
\shortauthors{Huang et al.}
\begin{document}
\title{Infrared Luminous Lyman Break Galaxies: A Population that Bridges LBGs and SCUBA Galaxies}
\author{J.-S.~Huang,$\!$\altaffilmark{1}
D.~Rigopoulou, $\!$\altaffilmark{2}
S.~P.~Willner,$\!$\altaffilmark{1}
C.~Papovich,$\!$\altaffilmark{4}
C.~Shu, $\!$\altaffilmark{5,}$\!$\altaffilmark{6}
M. L. N. Ashby,$\!$\altaffilmark{1}
P.~Barmby,$\!$\altaffilmark{1}
K.~Bundy, $\!$\altaffilmark{7}
C.~Conselice, $\!$\altaffilmark{7}
E.~Egami,$\!$\altaffilmark{4}
P.~G.~P\'erez-Gonz\'alez, $\!$\altaffilmark{4}
J.~L~ Rosenberg,$\!$\altaffilmark{1}
H.~A.~Smith,$\!$\altaffilmark{1}
G.~Wilson, $\!$\altaffilmark{3} and
G.~G.~Fazio$\!$\altaffilmark{1}
}

\altaffiltext{1}{Harvard-Smithsonian Center for Astrophysics, 60 Garden Street,
Cambridge, MA 02138}
\altaffiltext{2}{Department of Astrophysics, Oxford University, Keble Rd, Oxford, OX1 3RH, U.K.}
\altaffiltext{3}{Spitzer Science Center,
Caltech, 1200 E. California, Pasadena, CA 91125}
\altaffiltext{4}{Steward Observatory, University of Arizona, Tucson, AZ 85721}
\altaffiltext{5}{Joint Center for Astrophysics, Shanghai Normal University, Shanghai 200234, China}
\altaffiltext{6}{Shanghai Astronomical Observatory, Chinese Academy of Sciences, Shanghai 200030, China}
\altaffiltext{7}{California Institute of Technology, MS 105-24, 1201 E. California, Pasadena, CA 91125}


\begin{abstract}

A deep mid- and far-infrared survey in the Extended Groth Strip (EGS)
area gives 3.6 to 8\micron\ flux densities or upper limits 
for 253 Lyman
Break Galaxies (LBGs).  The LBGs are a diverse population but with
properties correlated with luminosity.  The LBGs
show a factor of 30 range in indicated stellar mass
and a factor of 10 in apparent dust content relative to stellar mass.
About 5\% of LBGs are luminous at all wavelengths with
powerful emission at rest 6\micron.
In the rest 0.9 to 2\micron\ spectral range 
these galaxies have stellar spectral slopes
with no sign of an AGN power law component,
suggesting that their emission is mainly powered by intensive star
formation.  Galaxies in this luminous population
share the infrared properties of cold SCUBA
sources: both are massive and
dusty starburst galaxies at $2<z<3$; their stellar mass 
is larger than $10^{11} M_{\odot}$. We suggest that these galaxies are
the progenitors of present-day giant elliptical galaxies, with a
substantial fraction of their stars already formed at $z \approx 3$.

\end{abstract}

\section{Introduction}

The two most efficient ways of identifying galaxies at high
redshifts are first, the Lyman break broadband-droput technique, 
and second, observations at submillimeter 
wavelengths. For example, surveys with the Submillimeter Common-User
Bolometer Array (SCUBA) on the James Clerk Maxwell Telescope 
reveal high-redshift objects via emission from the cold dust they
contain \citep{ivi02,sma02, cha03}.  
Conventional wisdom is that the submillimeter 
observations preferentially select starburst galaxies at $z>2$ because
the far-infrared luminosity peak shifts into the submillimeter
band \citep{ivi02, sma02, cha03}.  Most SCUBA sources are too faint in
the optical and near-infrared bands for spectroscopic identification,
but \citet{cha03} and \citet{sim04} 
have confirmed that most SCUBA sources in their sample are at $z>2$.  
Despite all the effort that has gone into SCUBA searches,
Lyman Break Galaxies (LBGs) 
still constitute by far
the largest well-studied rest-frame-UV selected 
galaxy samples at high redshift \citep{ste03}.

  The relation between LBGs and SCUBA galaxies remains unclear, but
recently \citet{cha05} have confirmed that some SCUBA galaxies
have the typical rest-frame UV colors as LBGs at $z\sim 2$\citep{ste04}.  
Understanding
both populations requires a direct comparison of properties, such
as stellar mass and dust emission. 
The four bands viewed by the Infrared Array Camera 
(IRAC) aboard Spitzer, from 3.6$\mu$m to 8.0$\mu$m, cover 
spectral energy distributions (SEDs) for galaxies at $2<z<3$
in the rest-frame near-infrared, where the
luminosities are good indicators of stellar mass 
\citep{col01, dro04, gla04, bun05}. The Multiband Imaging Photometer 
for SIRTF (MIPS) 24$\mu$m band is an ideal probe of the PAH and 
hot dust emission from LBGs and SCUBA galaxies. Preliminary studies
show that Spitzer can easily detect both LBGs and SCUBA galaxies 
\citep{bam04,ega04,hua04}.

This Letter presents the first study of a large LBG sample in the Extended
Groth Strip (EGS) region based on Spitzer observations carried out with
deep IRAC and MIPS photometry. We report discovering a new type of LBG, 
the Infrared-Luminous LBG (ILLBG) with strong infrared emission 
at rest-frame 6$\mu$m, whose properties are very similar to 
those of SCUBA galaxies.  ILLBGs are massive, 
dusty, and powered mainly by extremely rapid star formation.

\section{Spitzer Observations}

IRAC \citep{faz104, faz204} observations of the EGS were 
carried out in two epochs: 2004 January
and June, covering a $2^{\circ} \times 10'$ strip.
The MIPS observations were done in 2004 June in scanning mode.
Limiting flux densities (5$\sigma$) are
roughly 0.5$\mu$Jy at 3.6 and 4.5\micron, 2.7$\mu$Jy at
5.8 and 8.0\micron, and 60$\mu$Jy at 24$\mu$m. 
We use the same data reduction procedure for both IRAC and MIPS data as
\citet{hua04} and \citet{ega04}.

The LBG sample in the EGS area was taken from \citet{ste03}.
Among a total of 334 objects in the \citet{ste03} catalog,
188 are spectroscopically identified as galaxies at z$\sim$3, 3 are QSOs,
3 are AGN, and 7 are stars.  The remaining 133 objects are not identified.
In the photometric catalog 253 objects are in the area covered
by both IRAC and MIPS imaging.  Of these, 211 are detected at 3.6$\mu$m,
199 at 4.5$\mu$m, 53 at 5.8$\mu$m, and 44 at 8.0$\mu$m.
We detect 11 LBG counterparts in the MIPS 24$\mu$m image including
all 3 QSOs and 1 AGN identified by \citet{ste03}.\footnote{
Two additional LBGs may be detected at 24\micron, but there are
non-LBG galaxies within 3$''$ of them on the sky.  Given the angular
resolution of MIPS, we cannot be sure which objects are responsible
for the 24\micron\ flux, and to be conservative we do not count these
two as 24\micron\ LBG detections.}
The 24$\mu$m-detected counterparts are also detected at all shorter wavelengths
and are among the brightest sources in the sample.
We define the LBGs with 24 $\mu$m detections 
($f_{24 \mu m}>60 \mu Jy$) as Infrared-Luminous Lyman Break Galaxies
(ILLBGs). The 24$\mu$m limiting flux is equivalent to 
$L_{6.2\mu m}=6 \times 10^8 L_{\odot}$ $\micron^{-1}$
for galaxies at $z=3$, or $L_{ir}=3 \times 10^{10} L_{\odot}$ if we 
use the M82 SED to
convert the luminosity at 6.2$\mu$m to the total infrared luminosity.

We compare the LBG sample with a set of Lockman Hole area SCUBA
sources having secure identifications \citep{hua04,ega04}.\footnote{
Finding counterparts to SCUBA sources is difficult because there are
many faint optical and infrared sources within the SCUBA error
circle.  For reliable identifications, radio observations are needed
to give accurate positions.  Four SCUBA sources in the
EGS IRAC$+$MIPS area have good radio observations \citep{webb03}, but none of
the counterparts thereby identified has a known redshift.  In fact,
SEDs suggest
that at least two of the four are at $z<2$. Thus the EGS SCUBA
sources do not provide a useful comparison sample, and we compare the
EGS LBGs with the Lockman Hole SCUBA sources instead.} \citet{ega04}
classified these SCUBA sources into two types (cold and warm) based
on the dust temperature inferred from their SEDs.  The cold sources
exhibit the 1.6$\mu$m H$^-$ opacity minimum bump in the IRAC bands, a
stellar feature typically seen in galaxy SEDs.  The cold SCUBA
sources are well approximated by the SED of Arp 220, a cold dusty
starburst.  The warm sources are AGN-dominated with a power-law
continuum SED, similar to that of Mrk 231. There are four cold and
two warm sources in the Lockman Hole SCUBA sample \citep{ega04,
hua04}.

\section{Infrared Colors of LBGs}

The LBGs exhibit a much wider range of flux densities in the 
IRAC bands than in the K band.
Figure~1 shows that the range of 3.6$\mu$m flux densities for the LBGs
 spans 4 magnitudes 
compared to only 1.5 magnitudes for the range of
K-band flux density seen by \citet{sha01}. 
Figure~1 also shows a correlation between [3.6] and $R-[3.6]$ for LBGs.
This relation is partially due to a selection effect
(all LBGs are brighter than $R=25.5$ \citep{ste03}), but
there is an obvious absence of LBGs with blue colors 
at higher 3.6$\mu$m flux densities.
All ILLBGs (as defined above) are redder than $R-[3.6]=1.5$, and 
most have $R-[3.6]>2.4$.
Three ILLBGs are so
red that they are close to qualifying as extremely red objects
(ERO) ($R-[3.6]>4$,
Wilson et al. 2004). LBGs detected at 8$\mu$m are redder than other galaxies 
($\langle R-[3.6]\rangle =1.95$ versus 0.94 for LBGs detected at
3.6$\mu$m but not 8$\mu$m)
but not so red as the ILLBGs. All six Lockman Hole SCUBA sources have
red $R-[3.6]$ colors \citep{wil04},
comparable to those of the ILLBGs.

Figure~2 compares the observed LBG colors with those predicted by 
two simple stellar 
population synthesis models
(single burst and constant star formation). These two models 
correspond to two extreme
cases of a model having a exponentially decay star formation rate,
$\psi(t)=e^{-t/\tau}$, where $\tau=0$ is the single burst model and
$\tau=\infty$ is the constant star formation model. 
For any given $E(B-V)$, 
the constant star formation model predicts the 
reddest $R-[3.6]$ color for galaxies at $z\sim 3$.
A combination of the two models with varying amounts of 
dust attenuation can reproduce almost all the LBG colors. 
\citet{sha01}, \citet{pap01},
and \citet{lab05} argue that a model with constant star formation rate
fits the SEDs of near-infrared bright LBGs better. 
This model should  also 
be very close to the real star formation history for 
the 8$\mu$m bright LBGs and ILLBGs. As shown in Figure~2, $g'-R$
is more sensitive to dust extinction, and $R-[3.6]$ is sensitive to
both age and dust extinction for galaxies with constant star
formation history. A rough estimate of the extinction range
as $0.1<E(B-V)<0.4$ is consistent with 
what \citet{sha01} and \citet{pap01}
measured in their LBG samples.
The LBGs with 8$\mu$m detections  
tend to be older but little dustier than those not detected at 8\micron. The  
ILLBGs are more extreme, most very red in $g'-R$ and $R-[3.6]$,
some beyond the model prediction. \citet{sha01} also found that
the SEDs of LBGs with the reddest $R-K_s$ defied any simple model fitting
and argued that those are starburst galaxies with multi-component
morphologies, each with different colors. Such a red color could
also be due to a significant contribution  from an obscured AGN, but
we argue below that significant AGN emission is inconsistent with the 
$[4.5]-[8.0]$ colors.

\section{Origin of 24$\mu$m emission from ILLBGs}

The observed 24$\mu$m emission ($\approx$6$\mu$m rest wavelength)
from the ILLBGs can come from either the power law component of an AGN
(e.g., \citealt{elv94, ega04, ste05}) or from
warm dust heated by a starburst (e.g., \citealt{san96,rig99}).
For strong starbursts, the 6.2$\mu$m PAH
feature can enhance the observed flux by perhaps
a factor of 1.5 \citep{tra01,pee04}.  
As \citet{ivi04} point out, a
key diagnostic for distinguishing between starbursts and AGN is the
existence of a change of slope around rest 3--4\micron:  strong
AGN have nearly a constant slope from 2 to 10$\mu$m, whereas
starbursts show a distinct minimum in the 3--4\micron\ range.
Translated to an observational criterion, rest $J-K_s > 2$ implies an AGN
(\citealt{cut01}; cf. \citealt{ivi04} Fig.~3).

Figure~3 shows the $[8.0]-[24]$ vs $J-K_s$ color-color diagram for 
starburst-AGN separation suggested by \citet{ivi04}. 
QSOs and warm SCUBA sources exhibit rest-frame $J-K_s$ color as red as 
local AGNs \citep{cut01}, while all  ILLBGs except one 
have $J-K_s<1.5$. Therefore we suggest that
the 24$\mu$m flux densities of ILLBGs are mainly due to strong
dust emission caused by
intensive star formation. ILLBGs have much redder $[8.0]-[24]$ colors
than QSOs and AGNs, steeper than typical power-law SEDs but consistent
with strong starbursts.
The SCUBA sources occupy the same overall
portion of color-color space in this plot as do
ILLBGs, AGNs, and QSOs at $z \sim 3$. In other words,
the cold SCUBA sources appear to be Arp-220-like starburst galaxies 
at the upper-left
corners, and the warm SCUBA sources are AGNs.
We conclude that ILLBGs, like cold SCUBA sources, are starburst galaxies.

\section{Characterizing the Galaxy Population at $z=3$}

The faint blue LBGs that lack 24$\mu$m detections
could have dust emission with flux densities below our detection threshold.
For example, even blue dwarf galaxies are found to have dust emission 
at rest-frame 6$\mu$m, presumably  with PAH features \citep{ros05}.
We used a stacking technique to assess the average dust emission
for the blue LBG population at z=3.
The LBGs without 24$\mu$m detection were divided
into two groups for stacking: those with 8$\mu$m detections and those without.
There are a total of 198 LBGs lacking 24$\mu$m detections
including both spectroscopically identified and unidentified
ones.  In this subset, 31 LBGs have 8$\mu$m detections. 
The 8$\mu$m and 24$\mu$m images for both groups were stacked separately.

The stacking results show a significant difference between LBGs with
and without 8\micron\ detections.
The LBGs without 8$\mu$m detections have 
$\langle M_K(Vega)\rangle =-21.5$,
right in the range exhibited by local dwarf galaxies as shown in Figure~4. 
The upper limit in the  
$[8.0]-[24]$ colors is also consistent with colors of local blue dwarf 
galaxies. This strongly suggests that most faint blue LBGs in the sample 
have similar stellar masses, dust obscuration, and star-formation histories
as dwarf galaxies in the local universe. 
The LBGs with 8$\mu$m detections, including 
the ILLBGs, have $\langle M_K(Vega)\rangle=-24.5$, brighter than the local 
$M^*_K(Vega)=-23.4+5*\log(h)$ \citep{col01,hua03}. The 
$\langle M_K(Vega)\rangle$ 
and $\langle [8.0]-[24] \rangle$ for the 8$\mu$m-detected LBGs 
are only slightly fainter and bluer than those of ILLBGs,
suggesting that they may be similar kinds of galaxies but  a little fainter
and bluer, just below the 24$\mu$m limiting flux density. 

  All the results show that the LBG population is far from 
uniform. Figures~1 and 4
demonstrate that the LBGs with the largest stellar luminosities, as indicated
by 8$\mu$m flux densities, are older and dustier than 
the LBGs with lower stellar
luminosities. A larger LBG sample with infrared photometry and deeper
8\micron\ observations are needed  to 
tell whether there are  two distinct populations or a 
continuous range of galaxy properties.

\citet{rig05} conducted a detailed analysis of the stellar populations and 
estimated stellar masses for the 8$\mu$m-selected LBGs.  
All were found to be massive galaxies($>4\times 10^{10} M_{\odot}$) 
with the ILLBGs tending to have
higher masses ($> 10^{11} M_{\odot}$).
These masses are 
consistent with the mean dynamic mass of
$1.5 \times 10^{11} M_{\odot}$ for starburst-type SCUBA 
galaxies \citep{swi04, tec04}. 
Massive galaxies with $10^{11} M_{\odot}$ have also been found 
at $1 \la z \la  2$ \citep{cim02, dad04,sar04,
gla04, tec04, lab05}. These observations challenge current
theoretical predictions of galaxy
formation \citep{kof99,shu01,bau03, som04}. This study 
provides further confirmation
that at least some massive galaxies were formed by $z=3$, 
favoring galaxy formation and 
evolution models assuming more rapid stellar mass accumulation scenarios
\citep{nag05}.

\section{CONCLUSIONS}

This infrared study of LBGs reveals that they are a diverse 
group in terms of both their masses 
and dust contents. The observed 8$\mu$m (rest
2\micron) flux densities, which are roughly 
proportional to stellar mass, extend over at least 1.5 orders of
magnitude, and the ratio of 24 to 8$\mu$m flux densities, which
measure the fraction of ISM emission, extends over more than an order of
magnitude. 

 Among 253 LBGs in the EGS area covered by Spitzer, 5\% were detected with 
$f_{24\mu m}>60 \mu Jy$. We refer to these as ILLBG. 
ILLBGs have much redder observed $R-[3.6]$
colors than the other LBGs in the sample. The rest-frame 
$J-K_s$ colors for most ILLBGs are stellar with
no sign of an AGN power-law component. Therefore we suggest that 
the 24$\mu$m emission is dominated by dust heated in a massive
starburst, probably 
including some contribution from
the PAH feature at rest-frame 6.2$\mu$m. ILLBGs and cold SCUBA  
sources share the same infrared properties, suggesting that
they are closely related and may belong to the same population. If so, 
it ought to
be possible to detect submillimeter emission from ILLBGs. 
Both ILLBGs and cold SCUBA sources tend to be massive with 
a typical stellar mass of
$10^{11} M_{\odot}$. Such masses make them candidates for the progenitors
of present-day giant ellipticals.

In contrast to ILLBGs, most LBGs in the sample are only detected 
at 3.6 and 4.5$\mu$m and have faint mid-infrared luminosities
and blue $R-[3.6]$ colors. Their lower average 8$\mu$m flux density
measured with the stacked image implies that the faint LBGs have much less
stellar mass than ILLBGs, in a range similar to local dwarf galaxies.
An upper limit of $[8.0]-[24]$ color for average faint LBGs is
also consistent with dust emission of local dwarf galaxies.

\acknowledgements

JLR has received support from an NSF Astronomy and Astrophysics Postdoctoral 
Fellowship under grant AST-0302049.
This work is based on observations made with the Spitzer Space
Telescope, which is operated by the Jet Propulsion Laboratory,
California Institute of Technology under NASA contract 1407. Support
for this work was provided by NASA through Contract Number 1256790
issued by JPL.

\clearpage

\clearpage
\begin{figure}
\plotone{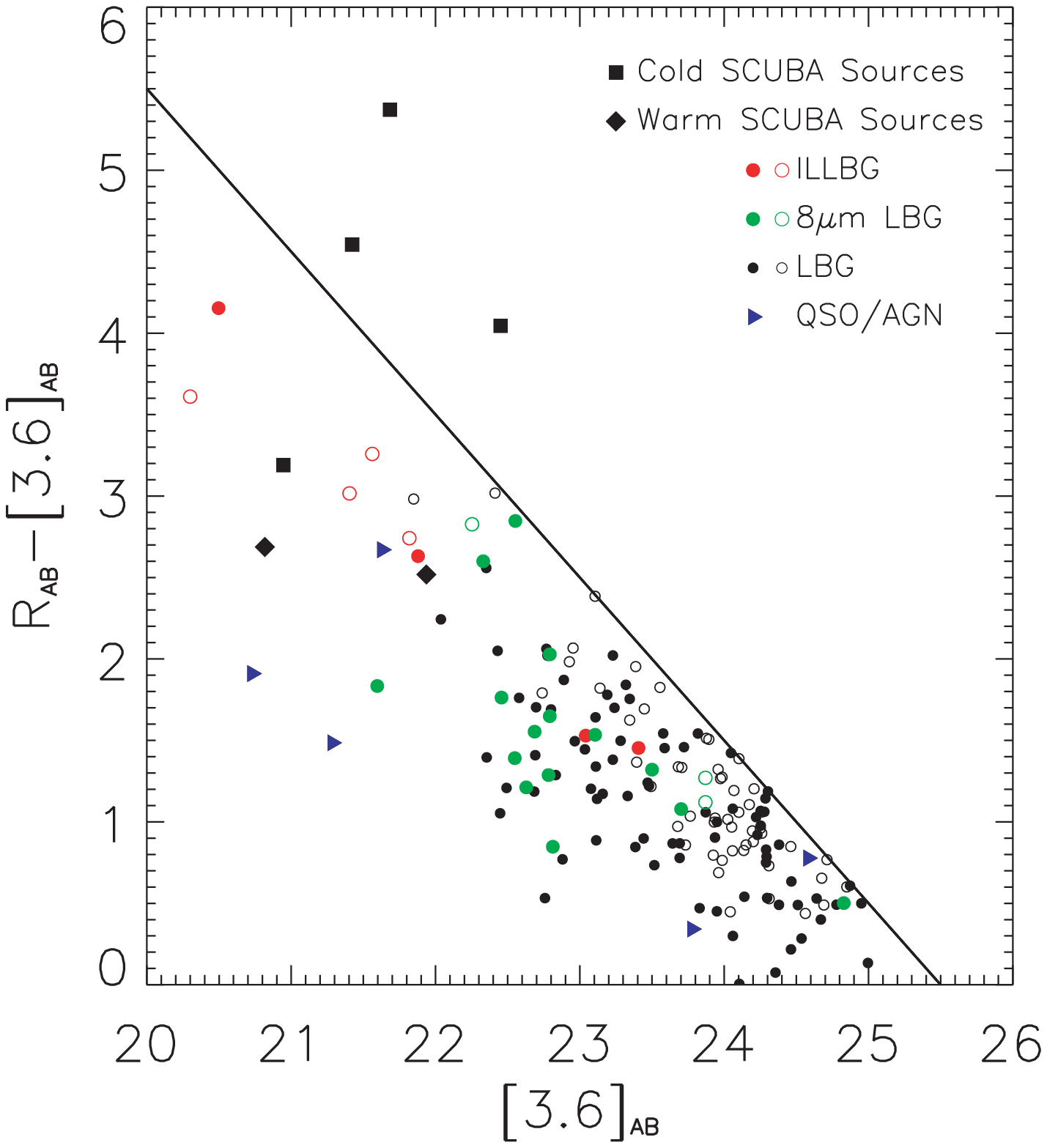}
\caption{Observed-frame $R-[3.6]$ versus [3.6] color-magnitude 
diagram for all LBGs
in our EGS sample.  
Black circles indicate LBGs detected only at 3.6$\mu$m
and 4.5$\mu$m.  
The green circles are LBGs detected at 8$\mu$m but not 24$\mu$m,
and red circles are LBGs detected in both 8$\mu$m and 24$\mu$m bands.
For all these objects, filled symbols designate sources having spectroscopic 
redshifts 
while open symbols are for objects with photometric redshifts only. 
Filled squares and diamonds represent the cold and warm SCUBA sources
in the Lockman Hole \citep{ega04}. The straight line is the R-band 
limiting magnitudes
of R=25.5 \citep{ste03}. 
\label{fig1}}
\end{figure}

\clearpage
\begin{figure}
\plotone{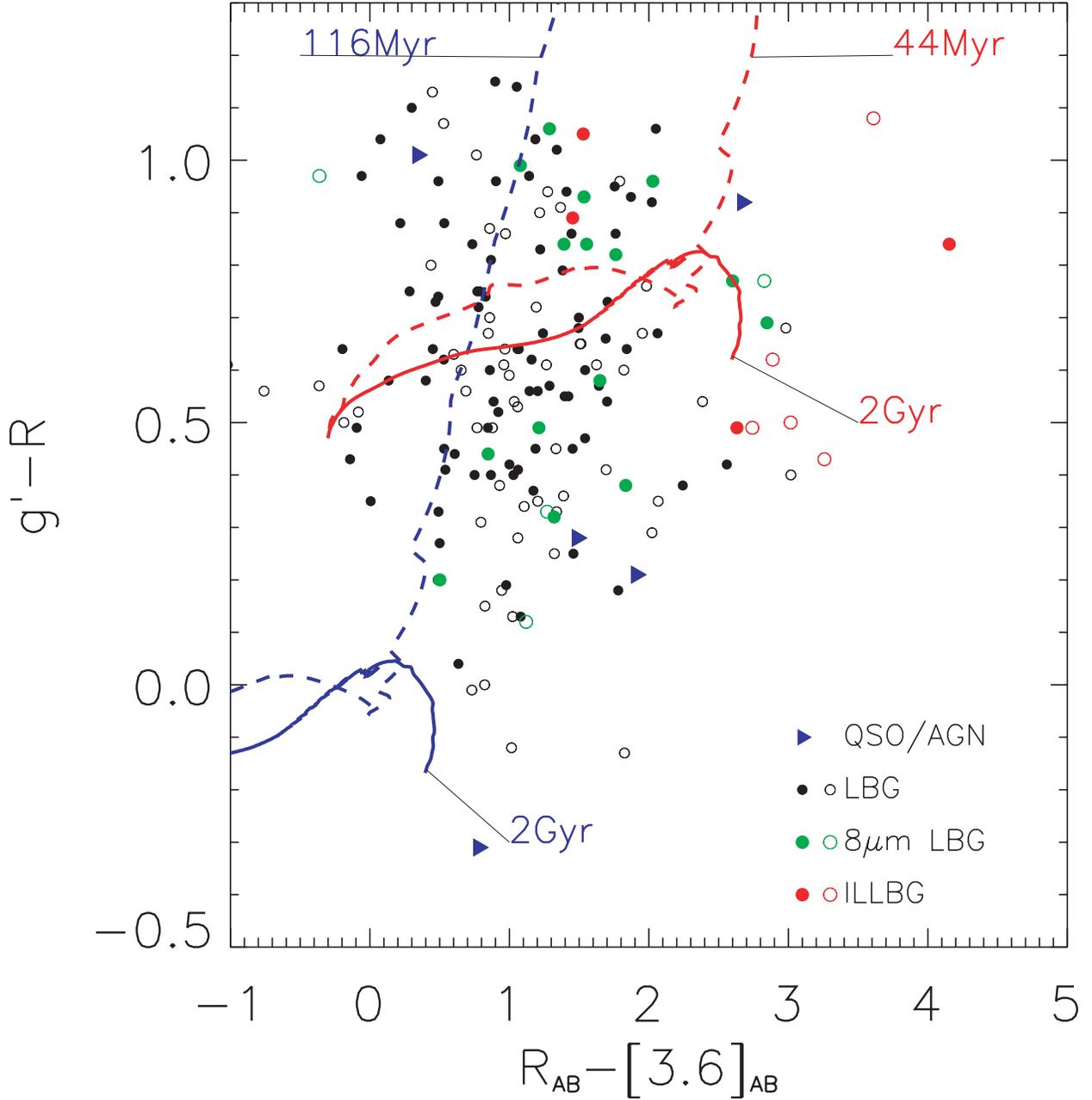}
\caption{Observed-frame $g'-R$ vs $R-[3.6]$ color-color diagram for all 
LBGs detected with IRAC. Symbols are
as in Figure~\ref{fig1}.  The lines show colors from stellar
synthesis models: dashed lines are the single burst models,
and solid lines are the constant star formation models. The blue lines
are dust-free models, and the red lines
are models with $E(B-V)=0.3$ and a \citet{cal00} reddening law.  
The models run from 0.1 Myr at the blue end of the lines for the reddened
models, off the plot for the dust-free models to 2 Gyr as indicated
for the constant star formation models, above the top of
the plot range for the single burst models. 
Maximum ages for the single burst models are
indicated near the top of the figure. Overall the set of models spans
nearly the entire color range of the data.
\label{fig2}}
\end{figure}

\clearpage

\begin{figure}
\plotone{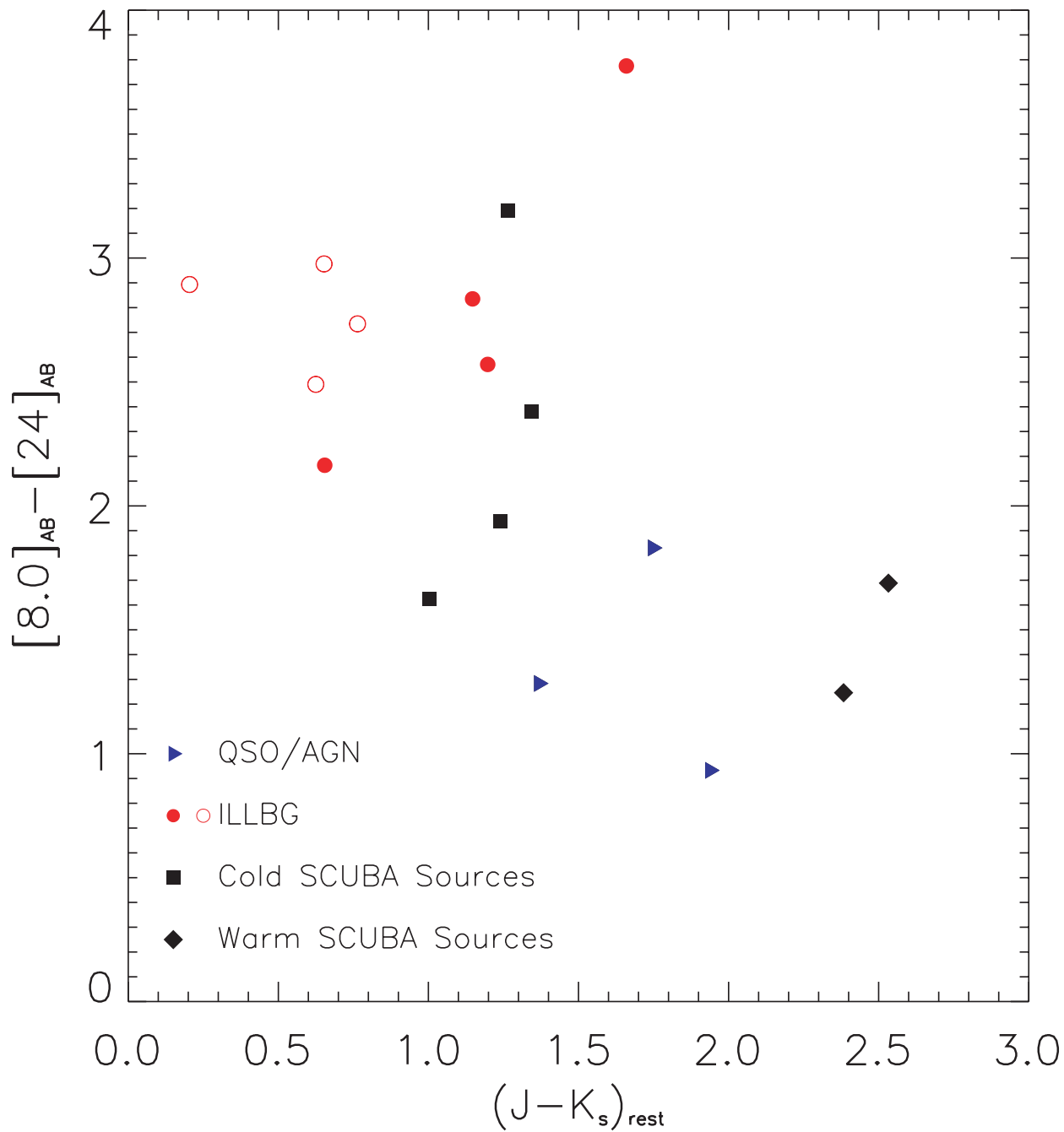}
\caption{Observed-frame $[8.0]-[24]$ vs rest-frame $J-K_s$
color-color diagram for ILLBGs, AGNs, 
and SCUBA sources. At z=3, observed $[4.5]-[8.0]$ is nearly equivalent to 
rest-frame $J-K_s$, and we have converted to $J-K_s$  using
K-corrections from \citet{hog02}.
The $J-K_s$ color is in Vega magnitudes for comparison
with the 2MASS results \citep{cut01,col01}, but $[8.0]-[24]$ is
in AB magnitudes.
\label{fig3}}
\end{figure}

\clearpage

\begin{figure}
\plotone{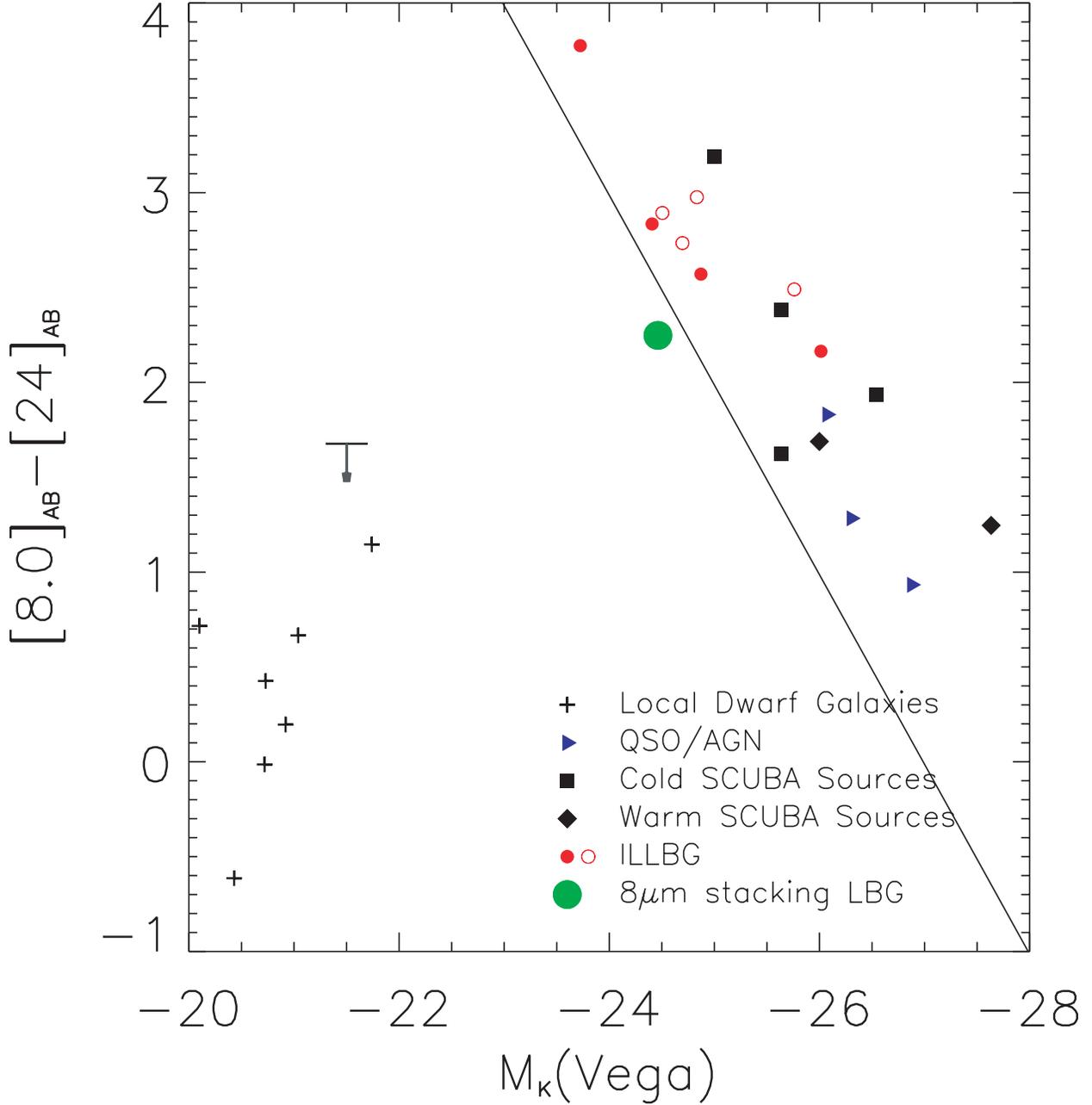}
\caption{Observed-frame $[8.0]-[24]$ vs rest-frame $M_K(Vega)$ color-magnitude
diagram for ILLBGs, AGNs, 
and SCUBA sources. 
We derived the observed-8$\mu$mto rest-K-band
K-correction with a spiral SED  for ILLBGs and cold SCUBA sources
and an AGN SED for AGNs, QSOs, and warm SCUBA sources.
A mean redshift of $z=2.97$ for LBGs was assigned to
those LBGs without spectroscopic redshifts.
The upper limit shown at $M_K(Vega)$=-21.5 
is for the stacked LBGs lacking 8$\mu$m detections.
The green filled circle is for the stacked LBGs having 8$\mu$m detections.
$[8.0]-[24]$ color for objects at $z \sim 3$ is equivalent  
to the rest-frame color $K-[6.0]$. We plot the  $K-[5.8]$
color for local blue dwarf galaxies \citep{ros05} for comparison. 
The solid line is  the MIPS 24$\mu$m limiting flux density for 
galaxies at $z=3$.
\label{fig4}}
\end{figure}

\clearpage

\end{document}